# An Impacting Descent Probe for Europa and the other Galilean Moons of Jupiter

P. Wurz[1,*], D. Lasi[1], N. Thomas[1], D. Piazza[1], A. Galli[1], M. Jutzi[1], S. Barabash[2], M. Wieser[2], W. Magnes[3], H. Lammer[3], U. Auster[4], L.I. Gurvits[5,6], and W. Hajdas[7]

1) Physikalisches Institut, University of Bern, Bern, Switzerland,
2) Swedish Institute of Space Physics, Kiruna, Sweden,
3) Space Research Institute, Austrian Academy of Sciences, Graz, Austria,
4) Institut f. Geophysik u. Extraterrestrische Physik, Technische Universität, Braunschweig, Germany,
5) Joint Institute for VLBI ERIC, Dwingelo, The Netherlands,
6) Department of Astrodynamics and Space Missions, Delft University of Technology, The Netherlands
7) Paul Scherrer Institute, Villigen, Switzerland.

*) Corresponding author, peter.wurz@space.unibe.ch, Tel.: +41 31 631 44 26, FAX: +41 31 631 44 05

# Abstract

We present a study of an impacting descent probe that increases the science return of spacecraft orbiting or passing an atmosphere-less planetary bodies of the solar system, such as the Galilean moons of Jupiter. The descent probe is a carry-on small spacecraft (< 100 kg), to be deployed by the mother spacecraft, that brings itself onto a collisional trajectory with the targeted planetary body in a simple manner. A possible science payload includes instruments for surface imaging, characterisation of the neutral exosphere, and magnetic field and plasma measurement near the target body down to very low-altitudes (~1 km), during the probe's fast (~km/s) descent to the surface until impact. The science goals and the concept of operation are discussed with particular reference to Europa, including options for flying through water plumes and after-impact retrieval of very-low altitude science data. All in all, it is demonstrated how the descent probe has the potential to provide a high science return to a mission at a low extra level of complexity, engineering effort, and risk. This study builds upon earlier studies for a Callisto Descent Probe (CDP) for the former Europa-Jupiter System Mission (EJSM) of ESA and NASA, and extends them with a detailed assessment of a descent probe designed to be an additional science payload for the NASA Europa Mission.

# 1 Introduction

This paper discusses a model impacting descent probe that is designed to be an additional science payload for a spacecraft performing a flyby of a planetary object without an atmosphere, such as the NASA Europa Mission (http://solarsystem.nasa.gov/missions/europaflyby), which we take as model case. However, the applicability of the proposed mission design is not limited to Europa, and it can be adapted to the more general case of future missions to other atmosphere-less planetary bodies of solar system, including the other Galilean moons of Jupiter, moons of other planets, dwarf planets, and Kuiper belt objects.

The impacting descent probe is a small carry-on spacecraft that is supposed to be released by the main spacecraft before an Europa flyby, and brings itself on a collisional trajectory with the moon's surface, acquiring and transmitting science data during its descent until impact. It increases the science return of the mission by providing very-low altitude (down to ~1 km) imaging of the surface, determination of the neutral exosphere composition, measurements of the magnitude and orientation of the body's own (if any) or induced magnetic field, and quantification of plasma currents near the surface. In addition, the probe provides precious technical information in support of future Europa landing missions: topographic imaging at a spatial scale relevant for a meters-sized lander and direct characterisation of the near-surface radiation environment.

Building upon an earlier study of a Callisto Descent Probe (CDP) for the former Europa-Jupiter System Mission (EJSM) of ESA and NASA (Wurz et al. 2009), this study provides an expanded and tailored mission scenario for a Europa Descent Probe (EDP) that could have significantly increased the science return of NASA Europa Mission (Europa Study 2012 Report, 2012) with a unique balance between science impact and engineering effort in terms of resources budget (mass), complexity, risk, and Technology Readiness Level (TRL).

# 2 Background

The exploration of planetary bodies of the solar system by means of impacting probes has been performed since 1958, when at the beginning of the Moon race both the USA and the Soviet Union launched several impactors towards the Earth's satellite. The first six spacecraft launched by the Soviet Union towards the Moon between 1958 and 1959 were indeed impactors, and one of them – Luna 2 – became the first man-made object to reach the surface of the Moon on 14$^{th}$ September 1959 (Huntress et al., 2003). Also NASA had an extended and sophisticated impactors programme named Ranger, whose goal was to obtain the first close-up images of the lunar surface that were desperately needed for the planning of future Apollo landings. Nine impactors were launched between 1961 and 1965. The first fully accomplished mission – Ranger 7 – returned 4,300 images of the lunar surface in 17 minutes of descent at 2.6 km/s, down to an altitude of only 488 m, revealing features as small as 38 cm across (Forney et al. 1965, Huntress et al., 2003). Ranger 7 was the first successfully deployed sophisticated impactor designed to acquire data during a high-speed descent on an airless planetary body, until very low altitude before impact. In fact, it can be regarded as the first spacecraft to match the definition of impacting descent probe, as used in this paper.

Impacting spacecraft played an important role not only at the beginning of the space era, but also in more recent times. It is important at this point to distinguish between two categories of impactors: orbiters that are eventually disposed by surface impact at the end of its mission and incidentally provide useful science during descent or upon remote observation of their impact, and properly called impacting descent probes explicitly designed to perform science measurements during their high-velocity descent towards the surface.

To the category of spacecraft disposed by impact belong, for example, the European Space Agency's (ESA) SMART-1 spacecraft (Racca et al. 2002) that underwent a controlled impact on the Moon at a velocity of about 2 km/s, during which it obtained surface images until an altitude of 12 km by means of its star tracker camera. The impact flash was successfully observed remotely with telescopes from Earth (Camino et al. 2007). Another recent example is the Chinese National Space Agency's (CNSA) Chang'e 1, which ended its mission with a controlled impact on the Moon at a velocity of about 1.6 km/s. Also this mission returned low-altitude imaging down to < 36 km (Ziyuan et al. 2010), although the array push-broom CCD employed was optimised for the 200 km orbit, thus providing discontinuous surface coverage at the lower altitudes (Liu et al. 2012). The MESSENGER mission of NASA was terminated on 30 April 2015 by impacting the spacecraft on the surface with almost 4 km/s. The most recent example is the impact of the Rosetta spacecraft on the comet Churyumov-Gerasimenko on 30 September 2016 with a very low impact speed of 0.89 m/s.

To the category of properly called impacting descent probes belongs the Indian Space Research Organization's (ISRO) Moon Impact Probe (MIP) on board the Chandrayaan-1 mission (Ashok Kumar et al. 2009), which during its 25 minutes descent at 1.7 km/s performed both surface imaging and mass spectrometry measurements of the lunar exosphere that provided the first 'direct' evidence of the presence of water in the tenuous lunar exosphere (Sridharan et al. 2010a, 2010b, 2015).

Actually, one could distinguish a third group of impacting probes where the main, perhaps only science objective, is the impact of the probe on the planetary body itself, while the impact is observed by instrumentation on an accompanying spacecraft. Examples of such probes are NASA's Deep Impact mission to comet Tempel 1 (A'Hearn et al. 2005) and NASA's the Lunar Crater Observation and Sensing Satellite (LCROSS) mission (Colaprete et al. 2010).

Based on ESA-NASA bilateral discussions a contribution by ESA to the Europa Mission of NASA was considered. In late 2015 two concepts have been studied in the Concurrent Design Facility of ESA (http://www.esa.int/Our_Activities/Space_Engineering_Technology/CDF): CLEO/I, a small-satellite concept flying by Io, and CLEP, a penetrator targeting Europa. However, both concepts turned out to exceed the allocated 250 kg mass budget. CLEP – the penetrator option – besides weighing a prospected 308.8 kg (including 20% system margin), also required a significant change of the mission profile to achieve a v-infinity at the release of the Penetrator Delivery System as low as 1.68 km/s. Moreover, the overall TRL of the payload has been declared as rather low (2–3), thus requiring significant development steps (ESA Report CDF-154(E)). CLEO/I – the Io bound 'orbiter' – with a prospected weight of 266.8 kg (including 20% system margin) in the nominal case only slightly exceeded the allocated mass. Options that fit within 250 kg have been presented, although with a strong negative impact on science (i.e., Io hyperbolic flyby option, named: CLEO/I hyper; Europa flyby instead of Io, named: CLEO-E; ESA Report CDF-154(D)).

The Europa Descent Probe (EDP) described in this paper is specifically designed and optimised to achieve its science goals during the short and fast descent towards the surface of Europa, until impact. Therefore, it represents the next instance of impacting descent probes on the timeline that begun in the 1950s with the NASA Ranger missions and continued until present times, with the most recent successes constituted by the 2008 ISRO's Moon Impact Probe.

# 3   Science and Engineering Objectives of EDP

The EDP addresses two main themes: Europa science and future mission support. From a science standpoint, the following investigations are performed by the EDP: i) Magnetosphere-Moon Interactions; ii) Exosphere formation and composition; iii) Geochemistry; iv) Geology; v) Geophysics; and vi) Habitability and Astrobiology.

The EDP would extend the science performed by the NASA Europa Mission by allowing a very-low altitude (~0–25 km) in-situ (exosphere, ionosphere, and magnetic field) and remote sensing (imaging) investigations. On the one hand, the probe provides a local (i.e., along the probe descent trajectory) low-altitude data set that can be correlated with the higher-altitude investigations performed by the main spacecraft. On the other hand, the probe equips the mission with a capability to address science objectives that are not in reach of the main spacecraft. Key investigation of this kind are: in-situ measurements of exospheric species with a low scale height (i.e., heavy molecules) below the altitude of the main spacecraft by the mass spectrometer, very-low altitude ionospheric and magnetic field measurements by the plasma instrument and the magnetometer that could possibly detect the effects of an intrinsic or induced magnetic field (due to the subsurface Europa ocean) and closing magnetospheric currents, and finally surface imaging at a resolution potentially higher than the one achievable by the spacecraft instruments. Besides, the probe could cover regions with high resolution imagery that not covered by the main spacecraft during its 45 flybys, since the probe impacts at the antipode of the flyby C/A sub-satellite point.

From an engineering standpoint, the EDP addresses two key themes in support of a future landing mission: surface roughness and radiation environment. Specifically, the probe's Wide-Angle Camera (WAC) provides the first-ever imaging of Europa surface at a resolution (< 30 cm/px in the last images) that will enable to assess the ice surface roughness at a spatial scale that is relevant for a future meters-sized lander. Moreover, the probe is equipped with a Radiation Monitor (RAD) that will provide profiling of the radiation environment along the descent trajectory, allowing to verify the radiation environment that currently is mostly depending on modelling (Evans et al., 2013) as well as the model prediction on the near-surface reduction of penetrating radiation (Paranicas et al., 2007). Given that radiation-shielding mass would be a significant fraction of the total mass of an Europa lander and thus a design driver, an in-situ characterisation of the near-surface radiation environment is important to validate the current models and to avoid the need for excessive design margins due to the uncertain radiation situation at the surface, which would result in a heavier lander.

Because of the dual nature of EDP addressing science and engineering goals, ideally the probe trajectory and impact ellipse are selected to target areas, such as Thera Macula, that have already been pointed out as likely future potential future landing sites (Pappalardo et al., 2013).

## 3.1 Europa's Exosphere

The EDP will address the chemical composition and density profiles of Europa's exosphere down to ~1 km by the Exosphere Mass Spectrometer (EMS), thus providing a yet more solid ground to answer the key questions of the mission. Very little is currently known regarding the chemical composition of Europa's exosphere (McGrath et al. 2009). Oxygen atom UV emissions at 1304 Å and 1356 Å were detected in Europa's exosphere and from which a column density of $N_C(O) < 2 \cdot 10^{14}$ cm$^{-2}$ was derived (Hall et al. 1995). It has been inferred that the observed O atoms are the result of electron impact dissociation of $O_2$ and derived a disk averaged vertical column density of $N_C(O_2) = (2.4 - 14) \cdot 10^{14}$ $O_2$ / cm$^2$ for the trailing hemisphere, assuming both a uniform atmosphere and a spatially uniform electron impact excitation rate (Hall et al. 1995, 1998). These emissions are spatially inhomogeneous (McGrath et al., 2004). Moreover, a Na atmosphere of Europa has been discovered with a total mass of Na in the exosphere of about 840 kg, assuming spherical symmetry (Brown and Hill, 1996). Na and K were measured (Brown, 2001) in the extended exosphere of Europa, at distances of 6 – 13 $R_E$, with transverse column densities of $N_C(Na) = (1.5 - 3.7) \cdot 10^9$ cm$^{-2}$ and $N_C(K) = (4.5 - 13) \cdot 10^7$ cm$^{-2}$, and a Na/K ratio in Europa's extended atmosphere of 25±2. Recently, Roth et al. (2017) reported HST observations of the H corona at Europa from which they derived maximum exospheric densities at the surface in the range of $(1.5–2.2) \cdot 10^3$ cm$^{-3}$, confirming the abundances predicted by Monte Carlo simulations (Smith and Marconi 2006; Wurz et al., 2014). A recent review of the chemical composition and densities profiles for the relevant release processes was given by Vorburger and Wurz (2017).

The density profiles are calculated for species to be present in Europa's exosphere, either known from previous observations or expected and according to modelling and shown in Figure 1a and Figure 1b, respectively (Wurz et al., 2014), together with the limit of detection of the mass spectrometer, EMS (in orange) and the radiation-induced background near Europa (in yellow). For the calculation of the radiation background, the fluxes of energetic particles near Europa are considered, but not any geometric radiation shielding provided by the moon (Paranicas et al., 2007). Most of the energetic electrons impact at the trailing hemisphere of Europa; hence, fluxes of energetic electrons are about a factor 100 lower at the leading hemisphere (Paranicas et al., 2001).

Figure 1a presents the density profiles for exospheric species that are known to be present in Europa's exosphere (Wurz et al., 2014), Figure 1b shows expected species based on a Jupiter satellite formation model that has been presented in Vorburger et al. (2015). For the latter set of species a clear advantage is gained by getting closer to Europa by increasing the signal by a factor ~10. Moreover, heavier molecules, such as hydrocarbons, have a scale height that is significantly below even the closest C/A distance of the main spacecraft and their expected density is low, so they cannot be detected by instruments on the main spacecraft. However, these are the species of high interest to assess the habitability and possible presence of life. Below 10 km altitude even Kr and Xe are expected to become measurable, with noble gases being important tracers of the origin and evolution of planetary bodies. Note that the indicated radiation background is an upper limit since it did not consider any radiation shielding of the moon (Lasi et al., 2017). From Figure 1b it can be appreciated that EMS on EDP will be able not only to perform a better quantification of species that are already measurable at higher altitudes though with a very low S/N ratio, but also to detect heavier molecules of low abundance (such as hydrocarbons and noble gases) whose scale height is so low that their densities are too low to be detected at an altitude of 25 km.

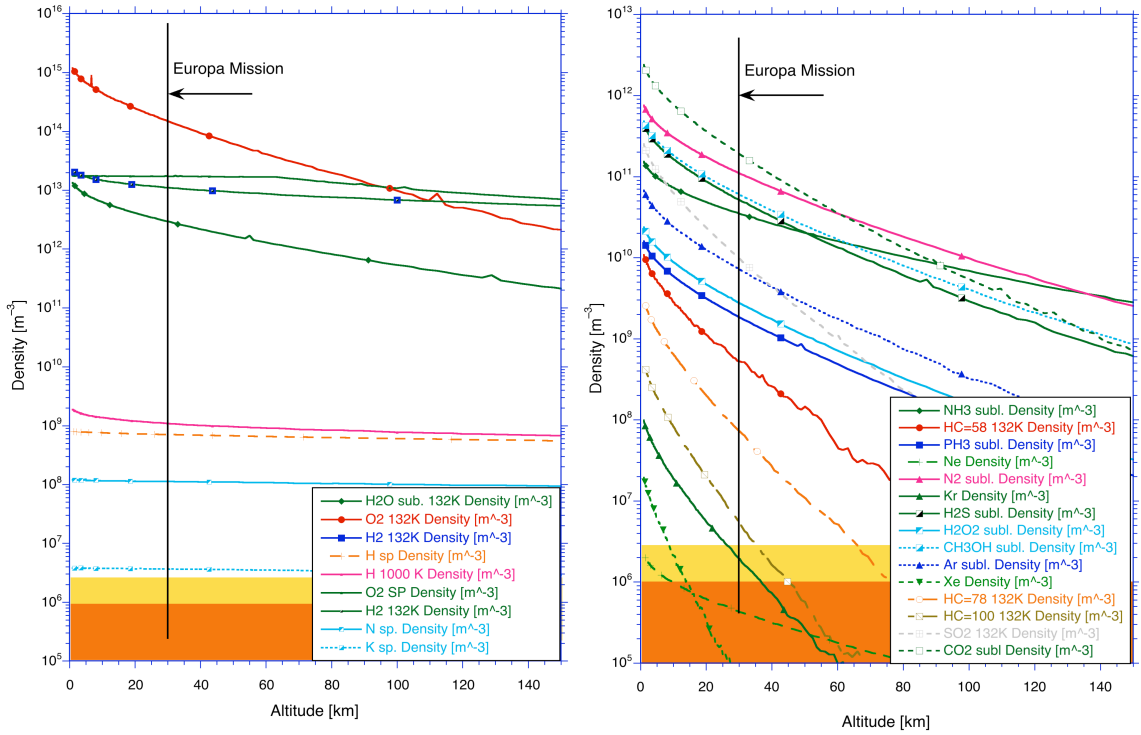

*Figure 1: Calculated exospheric density profiles for species known to exist in Europa's exosphere (a, left panel) and for species expected to be present based on the formation model (b, right panel). "SP" stands for sputtering, and "subl." stands for released together with sublimation water. HC are hydrocarbon molecules with the indicated mass. Intrinsic EMS background is given by red horizontal bar, the additional background from penetrating radiation is given by the yellow bar (Lasi et al., 2017).*

Besides providing detailed chemical composition of the exosphere along the probe's descent trajectory that is quantitatively and qualitatively better than the one provided by the main spacecraft (which in turn performs several flybys, thus covering larger regional area and different observation conditions), EMS also provides a profiling of these species' abundances that can be compared to the modelling predictions and correlated with local surface features that might be responsible for variations in the abundance of certain species and their processes of release from the surface.

It has to be noted that the radiation-induced background in Figure 1 does not take into account the reduction of the electrons fluxes of energy up to several tens MeV near Europa's surface (Paranicas et al., 2001), nor does it account for the geometrical shielding offered by Europa's solid angle as seen from EMS. Therefore, it is a worst case scenario and it can be confidently expected that EMS's detector will have a lower radiation-induced noise level that offers the capability to detect heavy species, such as noble gases and hydrocarbons, that according to the current conservative assumptions would be masked by the radiation-induced background.

So far $O_2$ has not been measured directly in Europa's exosphere, but has been inferred to be the dominant chemical species based on interpretations of the observed O I UV spectra (Hall et al. 1995, 1998; McGrath et al., 2004; Roth et al., 2016). However, the interpretation of Europa's oxygen emissions being principally due to electron impact dissociative excitation of $O_2$ has been challenged recently (Shemansky et al. 2014), where the authors argued that the observed O I UV emission is from electron excitation of atomic oxygen with a column density $4.7 \cdot 10^{12}$ cm$^{-2}$, i.e., approximately two orders of magnitude lower than the O$_2$ column densities derived in earlier studies (e.g. Hall et al. 1995, 1998). The existence of a dense O$_2$ exosphere and its extent is important for the interaction of the Jovian magnetospheric plasma with the exosphere, the associated mass loading of the plasma, the deflection of the plasma flow around Europa, and the shielding of the moon's surface from the magnetospheric plasma flow, see recent review by Plainaki et al. (2017). There are two populations of O$_2$ molecules in the exosphere, one from sputtering and radiolysis with a scale height of about 200 km and one from thermally accommodated O$_2$ with a scale height of about 20 km with much higher density (Vorburger and Wurz, 2017). Since the interaction of the exosphere with the magnetospheric plasma strongly depends on exospheric density and scale height it is important to directly measure the O$_2$ exosphere in situ, which needs a spacecraft approach Europa significantly closer than the scale height of the thermal exosphere. In addition, going all the way to the surface of Europa will show if the O$_2$ exosphere becomes dense enough so that collisions in the neutral exosphere become important as has been suggested (Shematovich et al. 2005; Smyth and Marconi 2006).

## 3.2 Plume Science

(Roth et al., 2014a) performed observations of ultraviolet emission lines of oxygen (at the 130.4 nm and 135.6 nm emission line) and hydrogen (Lyman-α) at Jupiter's moon Europa in December 2012 with the Hubble Space Telescope. From these observations the existence of water vapour plumes near the south pole was inferred, which persisted during the 7 h of observation time. From these observations the derived height of the plume is about 200 km and the inferred mass flux is 7000 kg/s with a gas temperature in the plume of 230 K.

The evidence of Europa plumes remains elusive since the plume was not seen in dedicated follow-up observation campaigns in 2014 and 2015 (Roth et al., 2014b, 2015, 2016). However, in recent far UV Hubble Space Telescope (HST) observations possible signatures of the plumes of this size have been identified in three out of ten measurements (Sparks et al., 2016). Huybrighs et al. (2016) showed that plumes that are a factor 1000 less massive than the ones observed with HST (Roth et al., 2004; Sparks et al., 2016), i.e., with a mass flux of about 7 kg/s, can be identified by the mass spectrometer and the plasma instrumentation on the JUICE mission during a flyby with 400 km C/A. Since the detection improves by being closer to the surface, even much smaller plumes would be identified by the EDP instruments during descent to the surface.

## 3.3 Magnetosphere – Moon interaction

Europa is embedded in Jupiter's magnetosphere where the corotating magnetospheric plasma is rapidly flowing past the moon and interacts electromagnetically with the moon's surface and its atmosphere. The interaction of the magnetospheric plasma with Europa's atmosphere, probably an exosphere down to the surface, is complicated because the ion gyro radii of the magnetospheric plasma are comparable with the scale lengths at Europa (Kivelson et al., 2009; Rubin et al. 2015) and the density and spatial extent of the exospheric species are based on modelling only, as has been reviewed recently (Plainaki et al. 2017). Since most of the exospheric species have scale heights of the order of 10 km, the plasma interaction with the exosphere will be close to the moon. Most of the atoms and molecules in Europa's exosphere are the result of sputtering, i.e., the impingement of plasma ions and energetic particles onto the surface releasing material from the surface into space (Johnson et al., 2009; Vorburger and Wurz, 2017). Thus the exosphere and the magnetospheric plasma are intimately coupled. Mass loading of the magnetospheric plasma by pickup of ionised exospheric species results in deflection of the plasma flow around Europa, in induced currents in the interaction region, pile-up of the magnetospheric magnetic field in front of Europa, resulting in a reduction of the magnetospheric plasma reaching the surface. These plasma-surface and plasma-exosphere interaction processes will all occur near the moon, at lengths scales commensurate with the scale height of the dominant exospheric species, thermal $O_2$.

These interaction processes are highly nonlinear, and also lead to the generation of various waves in the plasma (Kivelson et al., 2009), which will be observed in the magnetic field data close to their origin. Moreover, the measurement of the induced magnetic field will benefit from being close to the moon, and the magnitude and orientation of Europa's own magnetic field, admittedly unlikely to exist (Schilling 2006), may become feasible.

Since magnetic field and plasma measurements are in situ investigations any spacecraft has to travel to distances below the exospheric scale height to Europa's surface, ideally into the ionosphere, which is dangerous in a flyby but will be naturally accomplished by a descent probe.

## 3.4 Imaging

Obtaining observations at the highest optical resolution normally provides new insight into surface processes. Descent probes invariably give details that, although local by their nature, enhance the results from high resolution imagers onboard orbiters. For the Jovian system, imagers on descent probes can provide information on the properties of ices and their relationship to non-volatile species. Together with other payload elements providing compositional information, the images will provide context for many of these measurements allowing addressing questions in the area of geology, geochemistry, and geophysics.

The nominal Europa mission would include 45 flybys of Europa at flyby altitudes varying from 2700 km to 25 km (https://solarsystem.nasa.gov/missions/europaflyby/). The wide and narrow angle cameras of the Europa Imaging System (EIS) on the Europa mission would map most of Europa's surface at 50 m resolution, and would provide images of selected areas of Europa's surface at up to 0.5 m resolution. EPD imagery would clearly add to the inventory of high-resolution imagery of Europa. Moreover, EDP imaging of areas also covered by EIS high-resolution imaging will allow for imaging of areas on Europa at different phase angles, which gives a more complete picture of the moon's photometric behavior, local surface topography, and to understand what its surface is made of and how it varies from place to place.

## 3.5 Radiation

The radiation environment in Jupiter's magnetosphere is severe, with extremely high fluxes of energetic electrons and ions (Evans et al., 2013), which are even much more intense and with harder particle spectra than the terrestrial radiation belts. Models of the radiation environment have been formulated based on Pioneer 10 and Pioneer 11 data, with later additions of the Voyager 1 and 2 data, and the Galileo data, which is by far the largest data set. At present, these models interpolate over large volumes, since the data set is by far not complete (Garrett and Evans, 2015). Moreover, the fluxes of energetic particles are time variable, with fluxes changing by an order of magnitude at Europa (Garrett and Evans, 2015). In view of close flybys by a spacecraft and more so for a landed spacecraft on Europa's surface the knowledge of fluxes of penetrating radiation near and at the surface are of great importance for the design of the radiation shielding of the spacecraft and the understanding and forecasting of the radiation induced background in the detectors (Lasi et al., 2017). EDP measurements of the penetrating radiation near and at Europa will allow to verify the model predictions on the near-surface reduction of hundreds of keV to tens of MeV electrons (i.e., the dominating contributors to flux and dose on Europa) near Europa's surface (Paranicas et al., 2007), besides providing a full characterization of the radiation environment including fluxes of protons, gammas, and neutrons.

Moreover, such large fluxes of energetic particles hitting Europa's surface will create a lot of secondary particles being released from the surface, mostly gamma radiation, but also some energetic electrons and perhaps even neutrons. To characterise also these secondary radiation is one of the scientific objectives of the radiation measurements on EDP.

Finally, knowing the flux of penetrating radiation onto the surface will allow for estimating the changes of chemical composition of the surface due to radiolysis, with some of these products to be recorded by EMS.

## 3.6 EDP Impact Science

Using a shock physics code (Jutzi and Michel, 2004; Jutzi, 2015) we performed a set of calculations to estimate the effects of the impact of the descent probe on the ice surface of Europa. This shock physics code is a Smooth Particle Hydrodynamics (SPH) impact code specially written to model impacts on asteroids, which means it can handle different porosities of the planetary object as well as porous regolith surfaces.

Figure 2 shows the calculated impact plume for two impact angles, at 5° and at 18° for an ice surface with different material properties ("solid ice" and "regolith"). In the shallow impact, the projectile bounces at the surface and only a small impact plume is produced. In this case the plume size is very sensitive to the surface mechanical properties. For the steeper impact, the ejecta plume extends to more than several tens km, which is comparable to the plumes observed at Enceladus and should be easily observed with the remote sensing instrumentation on the main spacecraft in the backlight configuration. Although the nominal trajectory has been calculated for an impact angle of 18˚, the impact angle may be different due to the uncertainty in the actual realised descent trajectory, i.e., the size of the impact ellipse, which is preferably elongated towards lower angle impact, which maximise the probe flight time at lower altitude.

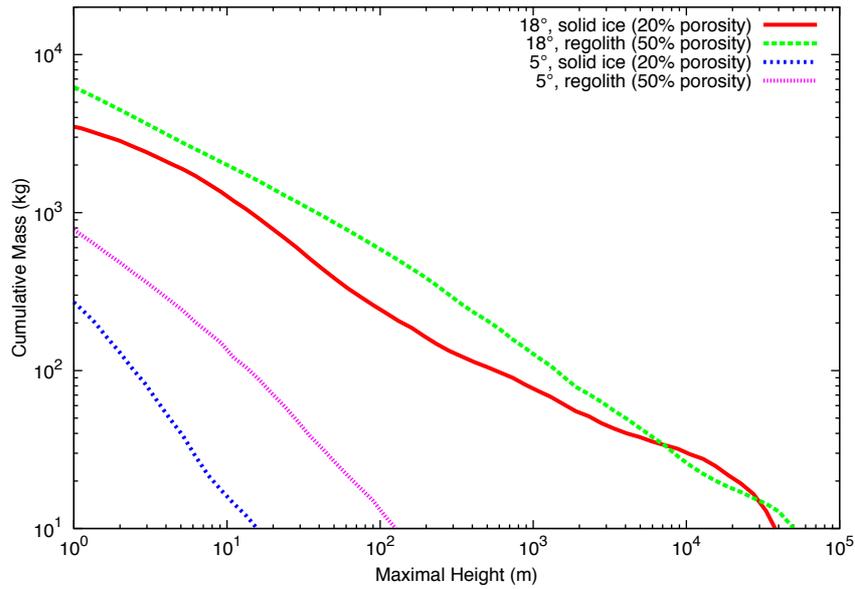

*Figure 2: Simulation of the impact of a 70 kg aluminium projectile at 4 km/s on Europa's surface using different impact angles (5° and 18°) and material properties ("solid ice" and "regolith"). Plotted is the cumulative mass of material ejected above a given maximal height.*

Finally, imaging of the EPD impact crater during consecutive flybys of the Europa spacecraft will allow to obtain the geometrical dimensions of the produced crater, its shape, which gives information on the mechanical structure of the surface. It also allows for a glimpse at the sub-surface material, for example, if the surface contains a lot of loose material (ice regolith), or, the other extreme, if it is made of solid ice. Using the scaling-law code from Holsapple (1993) we estimate that the EDP spacecraft of 70 kg impacting on an ice surface will produce a crater of about 10 m diameter, with a depth of 3 m, and an affected surface area of 30 m (average spall diameter). Such a surface feature should easily be observed in detail by the EIS imaging system.

# 4   Probe Design

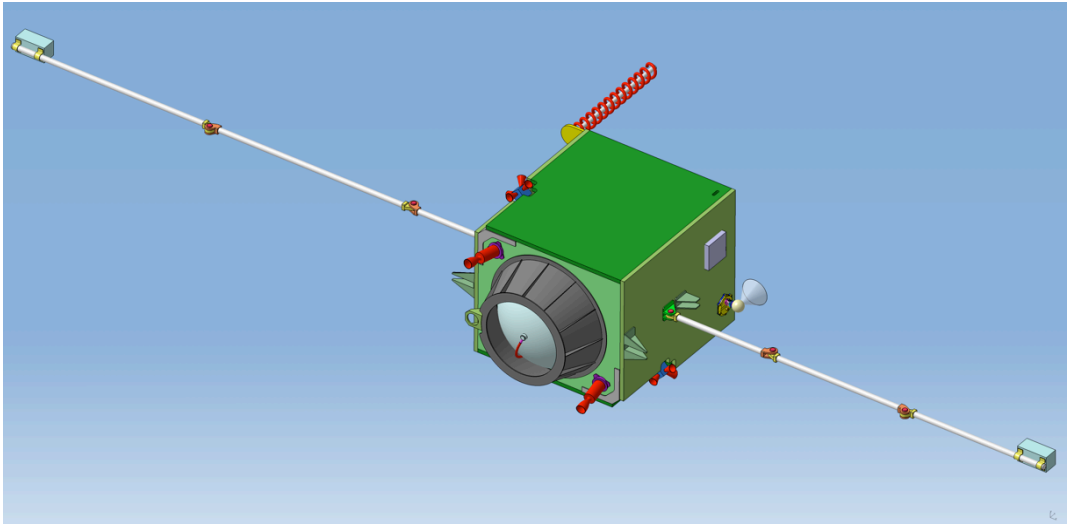

*Figure 3: CAD model of a possible probe implementation of the Europa Descent Probe based on the ALP-SAT satellite bus. The shape is approximately cubic, with long side being approximately 600 mm.*

The descent probe spacecraft design described in this paper is shown in Figure 3 and is based on ALP-SAT, a spin-stabilised microsatellite originally planned for a space weather monitoring mission, adapted to the Europa descent scenario also in consideration of some of the technical information contained in ESA's CLEO/P study reports. It has to be noted that the choice of the satellite bus should be subject to a more detailed analysis in Phase A of the mission, since other alternative buses may meet the mission needs (e.g., from Surrey Satellite Technology Ltd. or EADS CASA Espacio).

The EDP has an approximately cubic shape and externally hosts: the conical spacecraft adapter and separation mechanism, including power and communication interface for commissioning operation prior to separation, 2x20 N hydrazine thrusters acting as de-orbit engine, two pairs of 20 mN cold gas ($N_2$) thrusters for attitude control, one deployable helix-type Medium Gain Antenna (MGA) working in the UHF band (128–256 kbit/s, baseline: Proximity-1 Space Link protocol, CCSDS report 210.0-G-2), the camera, the mass spectrometer sensor head, the 3D plasma instrument, and the two deployable booms of the magnetometer. The attitude and orbit control system relies on a micro-star tracker, and gyroscopic and accelerometer sensors on a chip, with the optional use of the probe camera for probe-Europa distance estimation. Power is provided by Li-CF$_x$ primary batteries, which have a very high energy density (~670 Wh/kg after 8 years self-discharge, ESA Report CDF-154(E)).

The EDP is a spin-stabilised spacecraft. The spring loaded separation mechanism imparts an initial spinning momentum to the probe (~5 rpm), and the rate of spinning is regulated by the cold-gas thrusters between 40 rpm during de-orbit engine operation to achieve a delta-v of about 0.5 km/s and 15 rpm during the rest of the descent (Figure 4). Although the spin-stabilisation provides a simple method for stabilising the attitude of the probe, it imposes a number of constrains on the placement of the science instruments and the technical subsystems than a 3-axes stabilised spacecraft; however, the latter would require a significantly more complex system for attitude control. The major driver of the probe design are the science instruments and the technical subsystems with stringent Field of View (FoV) requirements, such as the camera, which needs to point towards the Europa surface, and the antenna, which needs to point to the relay spacecraft within a cone angle of ±14°. These subsystems are accommodated on the faces perpendicular to the spin axis, whose orientation is in turn determined by the relative position of the spacecraft and the moon's surface. In the mission scenario discussed in this paper, the probe and the main spacecraft are supposed to maintain continuous line-of-sight contact, in a geometry where Europa is placed slightly off the probe-spacecraft direction. Therefore, both the camera and antenna need to be on the probe's nadir face.

The EDP design shown in this paper represents only one of several possible configurations. If the camera and the antenna FoV point in the direction opposite to the de-orbit engine (as per Figure 3 configuration), the probe needs to perform a 180° flip manoeuvre after the de-orbit engine operation is completed to place both the antenna and the camera in the direction of Europa and the overtaking mother spacecraft. A different relative arrangement of the elements might save this extra manoeuvre, although any further optimisation of the accommodation of the various probe elements would require a detailed analysis of the actual mission profile.

The mother spacecraft uploads the flight sequence to the EDP before separation, through a hard-wired interface. After separation the probe operation is completely autonomous and communication is basically unidirectional with the uplink of science and housekeeping data from the probe to the mother spacecraft for relay. The mass and power budget of the EDP are shown in Table 1.

*Table 1 Mass and power budget of the EDP based on a modified ALP-SAT satellite bus. Mass budget includes mass needed for radiation shielding.*

| Subsystem | Mass [kg] | Power [W] |
|---|---|---|
| Science instruments total[1] | 19 | 36 |
| Structure including tanks | 10 | 0 |
| Primary batteries & power system | 10 | 0 |
| Propulsion and control system, incl. thrusters | 6 | 2 |
| Attitude and orbit control system | 2 | 3 |
| Hydrazine Thrusters (de-orbit engine) | 2 | 0 |
| Cold-gas ($N_2$) Thrusters | 2 | 0 |
| UHF communication system incl. antenna | 6 | 10 |
| Thermal control and hardware | 2 | 5 |
| On-board computer | 2 | 7 |
| Harness | 2 | 0 |
| Electronics common vault[3] | 10 | 0 |
| Hydrazine (de-orbit engine) | 17 | 0 |
| $N_2$ (Cold gas thruster) | 0.5 | 0 |
| Separation mechanism (on spacecraft) | 1 | 0 |
| **Total** | **91.5** | **63** |

[1]= *Details are given in Section 6.* [3]= *All electronics, including instruments', is hosted in a single common vault.*

Accounting for radiation shielding is of paramount importance for any mission bound to the Jovian system, because of the harsh radiation environment dominated by high-energetic electrons encountered in the Jupiter magnetosphere (Divine and Garrett, 1983; Cooper et al. 2001; Evans et al., 2013). Radiation shielding mass is considered in two different ways: i) for electronic components and other radiation sensitive components against total dose at the probe's end-of-life (dependent on the mission profile) and ii) for detectors against instantaneous fluxes of penetrating radiation causing background, which is dependent on worst-case flux during science measurement.

For the electronics shielding mass, it is assumed that the electronics, with the exception of few proximity elements, is hosted in a vault made of WCu, wherein the Total Ionising Dose (TID) is reduced below 100–300 krad (depending on the radiation hardness of the selected components). The vault approach has been successfully implemented in other missions, including NASA's JUNO (Stern, 2008) and ESA's JUICE (Grasset et al. 2013). Because the EDP is small, a single vault can accommodate all electronics, including science instruments', to minimise the surface to be shielded and therefore the mass.

The high instantaneous fluxes of penetrating radiation at Europa induce noise in the detectors, which is mitigated by dedicated shielding. The necessary shielding mass is estimated considering studies performed in the framework of ESA JUICE, such as GEANT 4 simulations and tests at high-energy electrons facilities (Tulej et al. 2015, 2016; Lasi et al. 2017). In this context, the small overall dimensions and mass of the EDP compared to a full-sized spacecraft is expected to be advantageous for instruments concerned with radiation background noise, such as cameras and particle instruments, because of the lower secondary radiation that are generated by the probe's nearby structures, such as gammas, which are difficult to shield.

It has to be noted that the sizing of radiation shielding against flux of penetrating radiation depends on the targeted Galilean moon and the sizing of shielding against total dose depends on the actual mission profile of the probe (e.g., the release time).

An early release of the probe by the spacecraft during its tour in the Jupiter system would reduce the probe's need for shielding mass against total dose and thus saves delta-v for the main spacecraft as well as probe. The current working hypothesis for the electronics vault is a 35 cm cube with 1 mm-thick WCu walls, which is sufficient to ensure a TID inside the vault in the Jupiter magnetosphere below about 100 krad as long as the probe is released before the spacecraft is exposed to ~100 Mrad of external Jupiter radiation. For comparison, it can be considered that the ESA JUICE mission is exposed to ~200 Mrad total dose in its three-years long Jovian tour. Therefore, the current assumed shielding mass should allow for sufficient flexibility on the probe release time, unless the release is planned for a very-late mission's phase. A very-late probe release is anyway not desirable for another reason: the earlier the probe is released from the main spacecraft, the lower the mass that the main spacecraft has to carry for the remainder of the mission, and the better the remaining fuel can be invested to optimise the science.

The choice of a different target than Europa, would expose the probe's detectors at higher (for Io) or lower (for Ganymede and Callisto) instantaneous radiation fluxes. A more detailed technical assessment is needed to quantify the change in instruments' detector shielding mass for mission scenarios different than Europa.

## 4.1 Planetary Protection

The EDP would be classified as Category IV in the definition of planetary protection by the COmmittee on SPAce Research, COSPAR (Kminek and Rummel, 2015); therefore, stringent planetary protection requirements apply. If the EDP will be the first mission to touch Europa surface, it will be very likely be required to abide to the highest level of safeguard. Additionally, an impacting probe is considered more critical than a landing spacecraft, because potential contaminants might end up beneath the surface and thus survive for longer time because the ice layer shields them from the radiation environment and its sterilizing effect. All in all, the mission will need to demonstrate that the probability of forward contamination of Europa is less than $10^{-4}$ per mission (National Research Council, 2000).

In a detailed evaluation of biological contamination risk and the required levels of biological cleanliness of the probe, it shall be considered that the probability of amino acids to survive an impact with a velocity of several km/s is on the order of $10^{-3}$ (Sugahara et al., 2014). These studies were done for impact on Earth, i.e., the impact plume was moderated by a substantial atmosphere. We expect that for an impact on an atmosphere-less body the heating of the impactor will be higher, since there is no atmosphere to absorb and dissipate energy, and thus the survival rate of amino acids and more complex molecules is expected to be even lower. In addition, the impact speed of the EDP is 4 km/s compared to a maximum of 1.6 km/s in the experiments performed by Sugahara et al. (2014), which will scale the energy density at the impact roughly with the square of the velocity and reduce the survival of complex molecules accordingly.

Finally, the required level of pre-launch cleanliness will have to consider also the bioload reduction during cruise and inside the Jovian radiation belt prior to release, and any potential for cross-contamination between the main spacecraft and the probe, if different cleanliness standards are applied. The combination of the above effects will determine the required level of pre-launch cleanliness of the EDP. Achieving those levels for EDP will be easier than for a larger spacecraft, because the whole probe might be sterilized at once: an approach that is considered unpractical for larger and more complex spacecraft and landers (National Research Council, 2000). The *whole-spacecraft sterilization* would reduce the complexity connected to sterilizing all individual components and subassemblies before integration.

# 5   Mission Profile

The design of the EDP mission profile was guided by creating the least possible impact on the main spacecraft. The EDP should be released from a regular flyby trajectory and bring itself onto a collision trajectory with the moon. Also, the science activities performed on the main spacecraft during the flyby should not be disturbed. The EDP activities should clearly be an addition to the main science and not drive the trajectory design nor the science operations of the main spacecraft.

The mission profile scenario calculated for the EDP descent trajectory is represented in Figure 4, which is assuming that the trajectories of the main spacecraft and the probe lie in the orbit plane of the moon. The concept of operation is described hereafter and is also summarised in Figure 5 using the Object-Process Methodology representation (Dori 2011; Crawley et al., 2015).

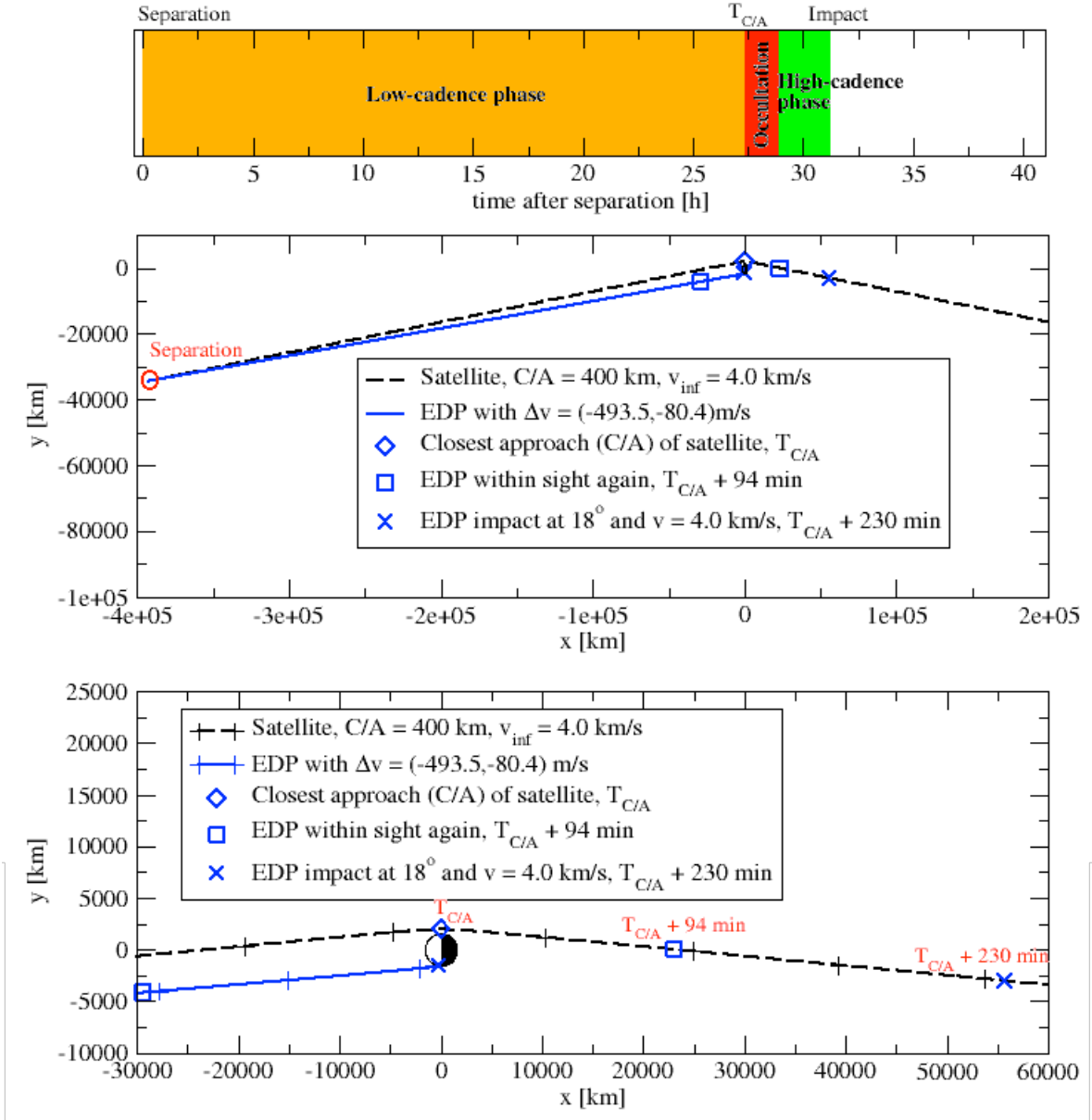

*Figure 4: Representation of the mission profile of the EDP. The top panel shows the length of the three mission phases (low-cadence phase from separation until C/A, occultation, and high-cadence phase from the end of occultation until impact), the middle panel shows the trajectories of the satellite (black dashed line) and of the EDP (blue line), the bottom panel is a zoom-in of the middle panel (here, the tick marks on the trajectories indicate time intervals of one hour).*

The trajectory analysed in this paper considers a Europa flyby with a closest approach (C/A) distance of 400 km and a flyby velocity of 4.0 km/s. The descent probe separates from the main spacecraft at a separation time of 27 h before the C/A of the main spacecraft with Europa ($T_{CA}$ – 27 h). Upon separation, the probe is oriented with the de-orbit engine towards Europa. After reaching a safe distance from the main spacecraft and a spinning rate of the probe of about 40 rpm, the de-orbit engine is operated to provide a sufficient delta-v to slow the probe down and place it on a hyperbolic collisional trajectory with Europa. The necessary delta-v depends on the details of the chosen descent trajectory. The de-orbit engine operation shall be completed within the first one hour from separation, allowing to perform a 180° flip manoeuvre that orients the FoV of the scientific instruments and the medium gain antenna in the ram direction. At the end of the de-orbit manoeuvre, the spin rate is reduced to 15 rpm, for compatibility with science instruments' operation requirements. From this point on the probe maintains the attitude until impact by means of the cold-gas thrusters without any further operation of its de-orbit engine to avoid interference with the science measurement, and the science phase begins.

Because the probe telemetry shall be relayed in real-time to the main spacecraft, it is necessary to maintain line-of-sight contact between the probe and the spacecraft during the descent. Two scenarios have been considered for the probe trajectory: one aiming at an impact location in the same region of the sub-satellite point at C/A (as baselined by Wurz et al. 2009) and one aiming at an impact at the antipodal point (Figure 4). The former requires that the probe is released by the spacecraft just few tens of minutes before closest approach, and requires a probe delta-v of ~0.9 km/s. This is necessary to avoid that the probe disappears under the horizon of the moon as visible from the spacecraft during the last, and most valuable, part of the descent. This scenario is not considered preferable because it would require the main spacecraft to sacrifice most of its science operation in proximity of C/A to support the probe's operations.

Impacting at the antipodal point, instead, allows the probe descent to happen when the main spacecraft is already past Europa by about 40 $R_E$, with full visibility of the probe low-altitude descent trajectory without interfering with spacecraft science during C/A. Actually, the probe is in eclipse between $T_{C/A}$ and $T_{C/A}$ + 94 min; therefore, the main spacecraft can perform its C/A science without any attitude constraint, and needs to re-establish contact with the probe only on the departing leg from Europa after C/A. The delta-v required for the probe is only 0.5 km/s in this case; most of it is needed to create a velocity difference between the spacecraft and the probe that ensures a proper timing for visibility of the probe from the spacecraft in the last 2 h of the probe descent. Because of these advantages, the antipodal impact scenario has been taken as baseline.

The trajectories shown in Figure 4 have been calculated for a v-infinity at the probe's release of 4.0 km/s and for a flyby altitude of the main spacecraft of 400 km. The assumptions on the trajectory of the mother spacecraft are not critical; similar EDP trajectories can easily be found for different C/A distances in the range from 25 km to 1500 km. Considering both the initial probe deceleration (delta-v) and the subsequent acceleration during descent by Europa's gravitational field, the probe eventually impacts the moon's surface at 4.0 km/s (incidentally numerically equal to the initial speed when released) and at an angle of 18° (impact ellipse analysis not performed). Of course, several other trajectories are possible, and a more detailed analysis should be performed to optimize science return as well as operational requirements. Ideally, the mission profile is such that the probe approaches Europa from the sun-illuminated side, with the impact occurring near the Sun terminator. This provides the best conditions for the probe surface imaging (~60° Sun phase angle) and provide optimal circumstances for the remote observation of the probe impact plume by the main spacecraft (see Section 3.6).

This trajectory solution has the additional advantage that the main spacecraft has completed the scientific activities related to its flyby, without disturbance from the probe, and only 94 min after C/A at > 15 $R_E$ it has to dedicate itself to the support of the EDP operations.

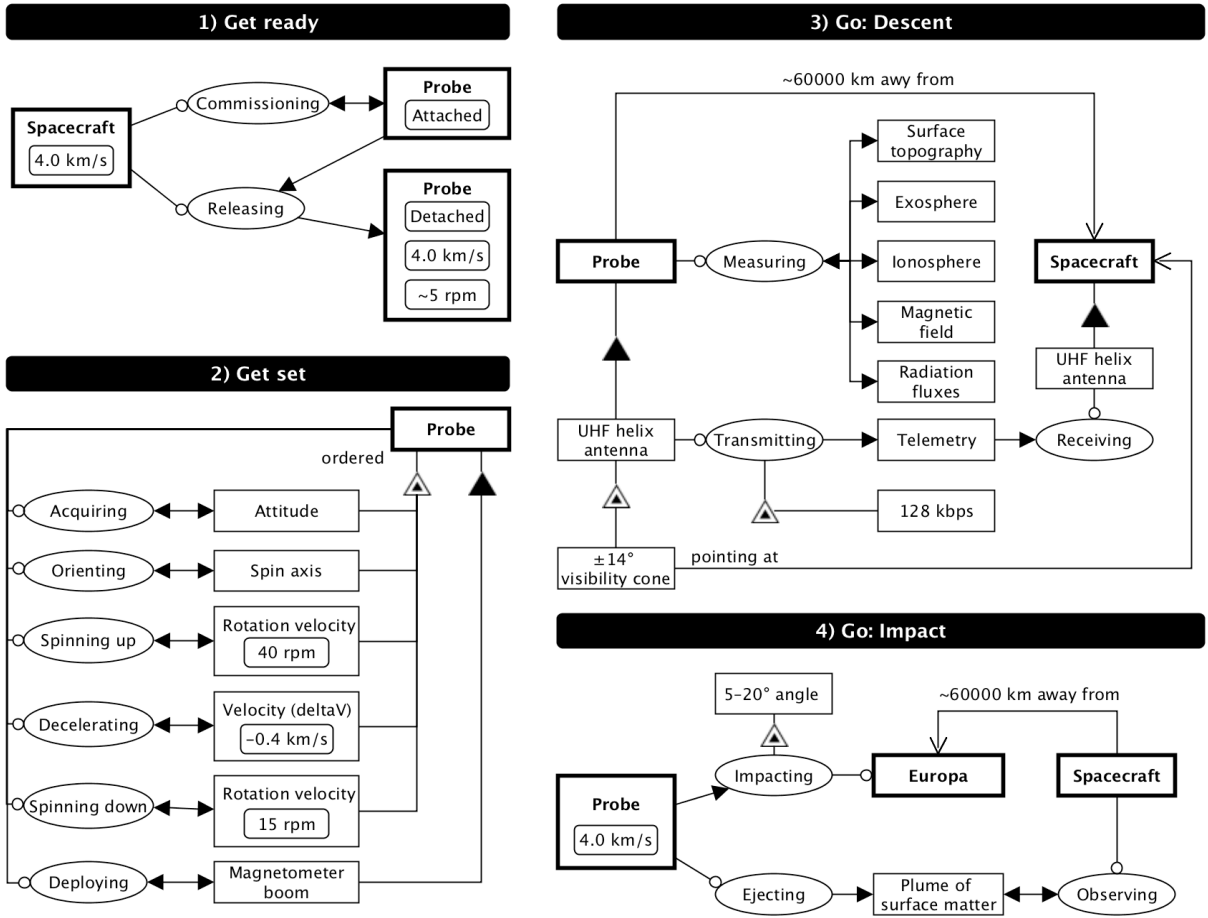

*Figure 5: Object Process Diagram representation of the EDP concept of operation, following the "get ready, get set, go" framework ("get unready" and "get unset" not shown here).*

## 5.1 Phases of the EDP Descent

The EDP descent can be divided into the three phases:
- *low-cadence* (high-altitude) observation, from separation to $T_{C/A}$
- *occultation*, from $T_{C/A}$ to $T_{C/A}$ + 94 min
- *high-cadence* (low-altitude) observation, from $T_{Science} = T_{C/A}$ + 94 min to $T_{impact} = T_{C/A}$ + 230 min

In the *low-cadence* phase, the probe's science instrument are performing observation at low-cadence (e.g., the mass spectrometer acquires background spectra with long 100–300 s integration time) and a low amount of telemetry is produced, thus requiring only a low-rate telemetry communication with the main spacecraft.

During the *occultation*, when Europa is between the probe and the main spacecraft, the probe enters an energy-saving mode, where only magnetic field and plasma data are collected at low data rate. The communication system is temporarily switched off and the data acquired during the occultation are stored locally and are uplinked after the occultation is finished.

In the final *high-cadence* phase, the probe science instruments are operating according to a pre-programmed sequence in their nominal science mode at their highest data rate, requiring the main spacecraft to maintain continuous communication with the probe. Since the probe science phase occurs at ~$T_{C/A}$ + 94 min to 230 min, when the spacecraft is already well past Europa at distances of 15 to 36 $R_E$, the spacecraft is able to follow the probe-related operations after its flyby's C/A science phase is mostly completed. This high cadence phase is the prime science phase of EDP ($T_{Science}$) and covers altitudes above Europa from 28,000 km (~18 $R_E$) all the way to impact (see Figure 4, bottom panel). The final descent phase with altitudes < 2 $R_E$ lasts 18 minutes.

The *high-cadence* observations ideally last all the way until the probe's impact; however, in consideration of the time needed to uplink the science data, there is a lower limit at which there is enough time before impact to upload the acquired data, depending on the size of a complete data product (e.g., image or spectrum). For instruments with a very low data product size (e.g., 500 bit per measurement for the magnetometer) all the data virtually until the impact can be uploaded to the spacecraft. For instruments with a high data product size (e.g., 90–180 kbit per spectrum of the mass spectrometer) the last retrievable data packet is acquired at about 1 km of altitude. Finally, for the camera both size of the image and motion blurring are a limiting factor, and the last acquirable good-quality image can be acquired at about ~ 4 km of altitude.

## 5.2 Europa plumes scenario

An alternative mission scenario would occur in case a Europa plume, as has been identified in December 2012 in UV images of the Hubble Space Telescope (Roth et al. 2014), would be detected prior to release of the EDP. The EDP mission profile could be adapted on short notice for the probe to fly through the plume, substituting the low-altitude science measurement with a *through-plumes* science phase. The plumes will be investigated by the mass spectrometer, by the plasma instruments, and imaged with the wide-angle camera. This will form a comprehensive data set to characterise these plumes in great detail.

Unfortunately, this mission scenario is hard to plan in advance; however, the science impact of the first-ever direct access of liquid water from Europa subsurface ocean would scientifically be enormously interesting in terms of habitability and extraterrestrial life investigation. Even if the main spacecraft might be equipped with even more capable science instrument for plumes investigation, letting the EDP flying through the plume first would be reduce any technical risk of flying the main spacecraft through an unknown environment likely to contain ice grains impacting on the spacecraft with 4 km/s, and for which it has not been necessarily designed for. Since the Europa mission plans 45 Europa flybys during the nominal mission to accomplish its science objectives, and probably considers even more flybys for a possible extended mission, the risk of flying through a plume would be much too high and should be passed on to EDP. Moreover, the flexibility in the release of the probe and the descent trajectory adjustment will add flexibility to access a plume that might be out of reach for the main spacecraft. The plume scenario is not discussed further this paper, but specific instrument's capabilities that may provide through-plumes investigations are reported in Section 6.

## 5.3 Impact Scenario

The impact of the EDP on Europa's surface can be observed in real-time by the main spacecraft with the remote sensing instrumentation from a distance of about 55 $R_E$ in a backlit configuration (see Figure 4), which is most favourable for imaging of the plume. The observation of the impact is an opportunity to measure the spatial extent, the mass, and the composition of the plume allowing for some insight on the surface composition and its mechanical properties at the impact location. Previous missions proved that valuable insights can arise from the remote observation of impacts with icy bodies. For example, NASA's Deep Impact mission to comet Tempel 1 (A'Hearn et al. 2005) deployed a 370 kg impactor at 10.3 km/s, which provided the opportunity of investigating both the chemical and physical properties of the ejected materials with a combination of space and Earth-based observation. More recently, NASA performed a controlled impact experiment into Cabeus crater of the Moon, the Lunar Crater Observation and Sensing Satellite (LCROSS) mission (Colaprete et al. 2010). Cabeus is one of the coldest craters on the Moon, where the 2300 kg Centaur stage of the Atlas V launch vehicle hit the surface at 2.5 km/s. The released plume was studied in detail with the scientific instrumentation on LCROSS Shepherding Spacecraft.

## 5.4 Post-Impact Scenario

The impacting descent probe described in this study is supposed to achieve all its science goals up to impact on the surface, and any residual post-impact functionality is considered only as optional add-on. However, since the early studies for a Callisto Descent Probe (Wurz et al. 2009) a spherical survival capsule equipped with a radio beacon and batteries sufficient for two weeks of operation was foreseen as a potential additional payload. The beacon signal will be transmitted with nearly-omni-directional antenna at the X-band frequency (8.4 GHz) with a power of about 3 W. The technical feasibility of surviving an impact at 4.0 km/s velocity has to be assessed in more detail by future simulation and experiments. As demonstrated, e.g. by the Huygens Probe (Lebreton et al. 2005) and in a number of experiments with the Venus Express (Duev et al. 2012) and Mars Express (Duev et al. 2016), such the signal can be detected by a network of sensitive Earth-based radio telescopes combined in a Very Long Baseline Interferometry (VLBI) system. As a minimum, VLBI tracking of the beacon's signal can provide range-rate (Doppler) and lateral position measurements of the probe in the interests of celestial mechanics and planetary dynamics. The science value of such the measurements will be considerably enhanced if the probe's transmitter is equipped with a sufficiently stable on-board oscillator.

A yet more speculative possibility would be the addition of a non-volatile mass memory in the survival capsule, wherein the probe would store additional selected science data recorded in the moments preceding impact that cannot be uplinked to the main spacecraft during descent. By modulating the beacon signal, the survival capsule could transmit a very limited but precious amount of data collected during the very last moments of the descent.

# 6   Model Payload

The engineering details and the capabilities of the instruments composing the EDP model payload are summarised in Table 2, including the lowest altitude at which the last data are supposed to be acquired and transmitted, based on the mission profile described in Section 5.

*Table 2 Summary of the key characteristics of the model payload of the EDP (more details for each instrument are given in Section 6 below). The last three lines are relative to the lowest altitudes data point. (\*) = Includes the post-processing of the data, such as compression.*

|  | WAC | EMS | PIECE | MAG | RAD |
|---|---|---|---|---|---|
| **Measured quantity** | Visible light imaging (2 band 450–600 nm, 650–850 nm) | Neutral atoms and molecules (< 10 eV) Range: 1–1000 amu | Ions (10 eV – 15 keV) Range: 1 – 70 amu | Magnetic field vector (~ 10 pT / sqrt (Hz) at 200 Hz) | Electrons, protons, ions, gammas, neutrons |
| **Key instrument parameters** | Resolution: 100 µrad/px | Mass resolution: $M/\Delta M = 1100$ Sensitivity = 1 #/cm$^{-3}$ | Energy resolution: $E/\Delta E = 0.07$ | Resolution: 10 pT | Accuracy: ~10% per 1s measurement. |
| **Data product** | Monochrome image in 2 bands Up to 2k x 2k 12 bit grey scale | TOF histograms Integration time: 0.1 – 300 s | 2D (32, 64, 128 E-steps) per 0.25, 0.5, 1 s. 3D per 8 s | Magnetic field vector | Particle fluxes |
| **Minimum acquisition time\*** | 1 per second (70 µs exposure) | 0.2 s cadence (0.1 s integration) | 0.25 s for 2D (azimuth and energy) 2 s for 3D | 0.1 s | 1 s |
| **Size after compression [factor]** | 246 kbit [200] | 90–180 kbit [10] | 10 kbit [2] | 0.5 kbit [2.6] | 1 kbit [2] |
| **Lowest altitude data point** | ~4 km | ~1 km | ~1 km | ~0-1 km | ~0-1 km |

The EDP is equipped with a non-redundant Data Processing Unit (DPU) which, depending on the system architecture, may simply validate and sort the compressed data streams of the different instruments to prepare them for upload to the main spacecraft, or be the single place where data conversion, such as processing and compression, takes place.

The instruments are pre-calibrated and do not need to be operated before the release of the probe, although a few health-check controls and commissioning procedures (e.g., decontamination heaters operation) might be needed shortly before release. However, it has to be mentioned that an additional and optional capability can be envisaged when EDP's PIECE and EMS could be occasionally operated via power and communication interfaces by the main spacecraft during cruise or the early Jovian tour, to provide technical synergetic tasks with other mission's instruments, such as spacecraft outgassing products characterisation and in-flight cross calibration (e.g., plasma instruments).

The technical budgets of the EDP model payload are detailed in Table 3.

*Table 3 Mass and power budget for EDP model payload. The mass of the instruments includes their proximity electronics and all structural parts. The radiation shielding mass of detectors against instantaneous fluxes, when applicable, is included in the total mass. The shielding mass against TID is excluded, because it is already accounted by the common vault mass.*

| Instrument | Mass [kg] | Power [W] |
|---|---|---|
| **WAC** | 1 | 5 |
| **EMS Sensor** | 4 | 12 |
| **MAG incl. booms** | 2 | 2 |
| **PIECE** | 4 | 7 |
| **RAD** | 1 | 1 |
| **Common DPU** | 3 | 4 |
| **Science instruments total (incl. ~20% margin)** | **19** | **36** |

## 6.1 The Exosphere Mass Spectrometer

The Exosphere Mass Spectrometer (EMS) is a neutral-gas time-of-flight (TOF) mass spectrometer that strongly builds on the heritage of neutral gas mass spectrometer NIM of the PEP experiment on JUICE (part of the Particle Environment Package – PEP; Barabash et al., 2013), and on earlier designs for Luna-Resurs/NGMS (Wurz et al., 2012b) and Rosetta/RTOF (Scherer et al., 2006; Balsiger et al., 2007).

EMS performs neutral-gas measurements in the mass range 0–1000 amu, with an integration time typically of: 100–300 s (background measurement at altitudes > 100,000 km), 1–5 s (at altitudes below 2 $R_E$), and 0.1 s (last points before impact). EMS has a mass resolution M/ΔM of 1100 at 84 amu, an instantaneous dynamic range of 6 decades at 1 s accumulation time, and a sensitivity of $10^{-16}$ mbar, corresponding to about 1 atom or molecule per $cm^3$ (Wurz et al., 2012b). EMS has a 60x10° FoV; the probe's spinning does not pose a measurement challenge for EMS as long as the ram direction is in the FoV, but it rather favours the removal of any contaminant possibly accumulated during cruise by exposure of the external surfaces to sunlight in the early phase of the descent. The neutral and ionised exospheric thermal gas (<10 eV) enters the ion source, wherein neutral gas is ionised by electron impact and ions are captured, and periodically injected in the TOF ion optics by a high-voltage pulser operating at 10 kHz repetition rate. The ions are accelerated in the ion source, fly through a field-free drift path and are reflected by the integrated reflectron towards the MCP detector. The detector signal is digitised with an ADC (2 GSPS) to provide waveforms that are summed up in histograms (0.1–300 s integration time) in the main controller. The histograms are transferred to the main DPU, where they are compressed using a custom implementation of a SPHIT algorithm (http://www.cipr.rpi.edu/research/SPIHT/spiht0.html).

No element of the ion optics needs to be radiation shielded, except the MCP detector. The detector has a very compact design, and the radiation-induced background at Europa can be reduced to an acceptable level using a 1-10-1 Al/Ta/Al sandwich for a total mass of 1.5 kg. This design is the same as developed for PEP's NIM, for which the Europa flyby was also the sizing case. The detector and its radiation shielding was validated by testing at the Paul Scherer Institute (PSI) with a monochromatic electron beam of energy up to 345 MeV (Tulej et al., 2015, 2016, Lasi et al., 2017).

The whole EMS sensor weighs about 4 kg, including all its dedicated electronics and radiation shielding elements. The power consumption is 12 W, similarly to JUICE's NIM.

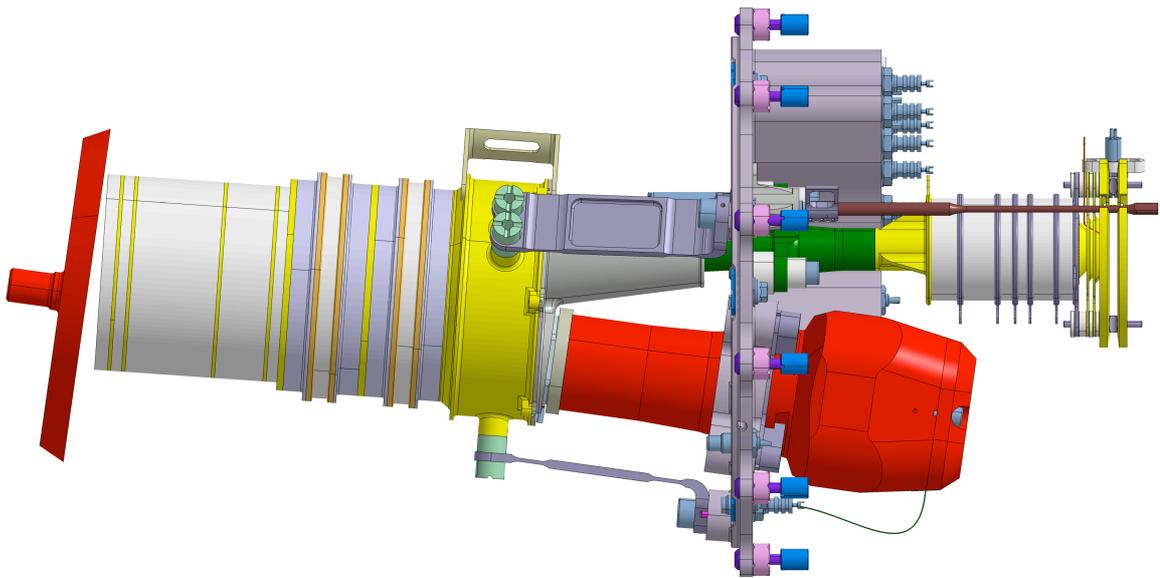

*Figure 6: CAD model of the EMS instrument. The elements for radiation shielding of the MCP detector are shown in red.*

## 6.2 Wide Angle Camera System (WAC)

There are several possible approaches to the development of a wide angle camera depending on the technical boundary conditions. Clearly, a descent camera system needs to obtain higher resolution data than the main spacecraft. For the purposes of this work we assume that the main spacecraft will obtain data from 25 km altitude at around 20 µrad/px scale leading to no better than 50 cm/px and worse if degradation through motion is present (based on the Europa mission of NASA). Note that the flyby speed of the Europa mission and the descent speed of EDP are about the same. Our target is therefore to obtain < 30 cm/px immediately before impact. Through a detailed calculation we have found that 4.3 s to acquire and transmit this final image is compatible with a reasonable instrument design. With an angle of attack of 10 deg with respect to the surface and a linear velocity of 4 km/s, we can compute the required angular scale and in combination with the choice of detector, we can set the focal length.

In general, it is best to start at the detector when constructing a camera design. We assume the SRI-designed complementary metal-oxide semiconductor (CMOS) 2k x 2k detector originally produced for the SoloHi project on ESA's Solar Orbiter. It has a 10 μm pixel pitch (e.g. Janesick et al., 2014). Hence, with an angular resolution requirement of 100 μrad/px this leads to an effective focal length of 100 mm. With the instrument pointed to the surface orthogonal to the velocity vector, the smear time (the time for the image to move one pixel in the focal plane) can be computed leading to maximum exposure times for 1 pixel of smear. The aperture and the filter bandpass can then be set to provide sufficient signal to noise. A higher resolution imager, possibly following the Halley Multicolour Camera concept, could be envisaged but the cost and complexity would be significant.

An F-number of 3.3 appears to satisfy the requirements giving a 30 mm aperture that could probably be manufactured from radiation hard glass. We would suggest use of a dichroic beamsplitter to obtain two different pass bands with approximately 150 nm bandwidth in the optical wavelength regime. Using standard values for the instrument transmission, we obtain around 30000 electrons within the smear time of 71 microseconds, 4 seconds before impact. This would be of the order of 1/3 of full-well for a typical CMOS detector giving an SNR of over 175.

The camera head should probably interface directly with the DPU of the probe itself and hence the mass and power of the instrument will be driven purely by the camera head itself. This can be extremely light and a mass of < 250 g should be easily achievable before radiation shielding. We estimate a further 750 g should be allocated for proximity shielding. The power requirement for the instrument itself will be low (~2 W) but the transfer of data across to the spacecraft will require power. A SpaceWire connection to the DPU could be foreseen which provides around 60 Mb/s (net) allowing full-frame transmission of a 2k x 2k detector in < 1 s. We allocated 3 W for this. The time available to compress data is almost negligible and hence if transmission to the mother spacecraft is a bottleneck, reduction in the size of the frame or on-chip binning might be necessary.

The rotation rate of the probe requirement is an interesting one. If the rotation rate is too slow, then there is a non-negligible chance that the imager will be pointed away from the surface shortly before impact. Hence, a faster rotation rate is foreseen with around 15 rpm (identical to Giotto for example) which would guarantee the surface being in view every four seconds. Optimisation of the image acquisition to avoid images of space could be envisaged if a reference pulse defining a complete rotation is implemented. Operation at slower spin rates guaranteeing observation of the surface in the final seconds would almost certainly require a movable mirror or multiple camera heads.

## 6.3 Plasma Ions and Electrons Close Europa (PIECE)

The Plasma Ions and Electrons Close to Europa (PIECE) instrument measures the 3D distribution function of positive ions, electrons, and negative ions. The instrument builds on the heritage of instruments flown on previous missions, including NPD and IMA of Aspera-3/4 on Mars and Venus Express (Barabash et al., 2004, 2007), SWIM of SARA on Chandrayaan-1 to the Moon (Barabash et al., 2009), and PRIMA (Prisma, in Earth orbit). Besides, a similar instrument is currently part of the science payload of the ESA JUICE mission, which has similar requirements both in terms of performance and in terms of radiation hardness than needed for the EDP.

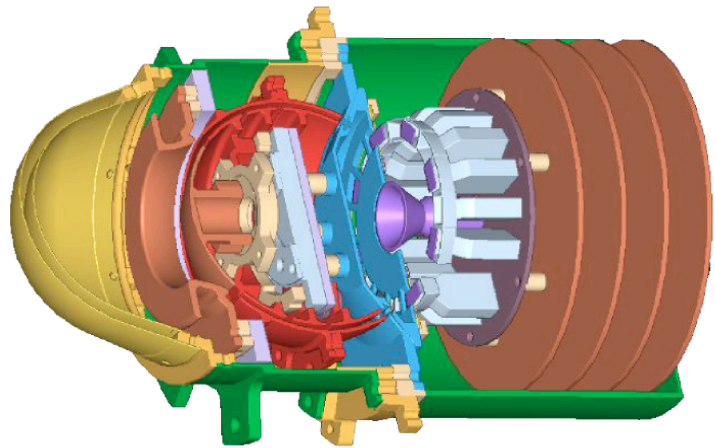

*Figure 7: Design drawing of the PIECE instrument, with the top-hat energy analyser on the left, followed by the time-of-flight section, the array of SEM detectors, and the electronics.*

PIECE is capable of measuring ions in the energy range 10 eV – 5 keV (or in the higher energy range ~50 eV – 25 keV, depending on the probe electrical potential) with high mass resolution, high time resolution, sensitivity, wide dynamical range, and capability to measure charge-state. The sensor includes a top-hat type electrostatic analyser (EA), and TOF cell based on surface reflection. Azimuth angle is measured instantaneously by imaging the arrival angle by the EA onto the image plane, and angular resolution is provided by 16 custom ceramic channel electron multipliers (CEMs). $4\pi$ angular coverage is achieved with a single FOV of $10° \times 360°$ together with the spacecraft spin. The particle energy is scanned by performing 16 energy steps during 0.25 s. Thus, a full 3D distribution is available after half of a spacecraft rotation, i.e., after 2 s. The CEMs collect the secondary electrons generated upon reflection of particles from 16 start surface areas arranged at the EA exit. These 16 signals also serve as start signals for the TOF measurement. A single CEM collects TOF stop signals from a centre stop surface. All CEMs are placed below the top-hat analyser, allowing use of the entrance and EA structure for radiation shielding. To improve radiation tolerance, rugged CEMs are used in the sensor head rather than MCPs (Wurz et al. 2009).

*Table 4: Capabilities of the PIECE instrument*

| Parameter | PIECE |
|---|---|
| Measured particles | Positive, negative ions, electrons |
| Energy range | 10 eV – 5 keV (low range) |
| | 50 eV – 25 keV (high range) |
| Resolution, $\Delta E/E$ | 20% |
| Mass range, amu | 1 – 70 |
| Masses resolution, $M/\Delta M$ | 2-3 (D-channel) |
| | 30 (C-channel) |
| FoV | 10° x 360° |
| Ang. Resolution | 10° x 22.5° |
| Time resolution | 2D per 250 ms |
| | 3D per 2 s |
| Dynamic range or G-factor | Total: $8.9 \cdot 10^{-3}$ cm$^2$ sr eV/eV |
| | Pixel: $5.6 \cdot 10^{-4}$ cm$^2$ sr eV/eV |

## 6.4 Magnetometer (MAG)

The Magnetometer (MAG) Instrument is a three axes fluxgate ringcore magnetometer. The MAG instrument consists of fluxgate sensor, front-end electronics, for each deployable boom mechanism. The sensors will be accommodated on the deployable booms (see Figure 3) to decrease the magnetic contamination of the magnetic measurements by other instruments and sub-systems on the EDP spacecraft. An example of a developed and tested boom with two fluxgate magnetometers is shown in Figure 8. The shown boom has been developed under ESA SOSMAG contract for service oriented magnetometer, and is to be used onboard KOMPSAT mission. MAG shall sample magnetic field vectors with 200 Hz. This corresponds to a spatial resolution of better than 20 m. The instrument capabilities are summarised in Table 5.

A similar instrument has been flown onboard the ESA mission ROSETTA (Glassmeier et al., 2007), the Japanese HAYABUSA mission (Herčík et al., 2016), and the NASA mission THEMIS (Auster et al., 2008). It will be launched onboard the MMO and MPO spacecraft of BepiColombo and is foreseen for the JUICE mission. The radiation tolerance is according the requirements of the JUICE mission. The proposed boom design is based on the heritage of PHILAE (Auster et al., 2007) and Venus-Express (Zhang et al., 2007), and on the development for a service oriented magnetometer system for the Korean KOMPSAT mission. Therefore the TRL level for all subunits is considered to be 7 or higher.

*Table 5 Capabilities of the MAG instrument.*

| Parameter | Value |
|---|---|
| Noise | ~ 10 pT / sqrt (Hz) at 1 Hz |
| Offset stability | < 2 nT / year |
|  | < 2 nT / 100 °C |
| Non-linearity | < $10^{-4}$ |
| Stability of magnetic axes | < 0.01° |
| Range | 5000 nT |
|  | to be in range with stowed boom |
| Resolution | 10 pT |
| Data acquisition rate | 200 Hz |
| Data transmission | 1.3 kbit / 0.1 s (before compression) |
|  | 0.5 kbit / 0.1 s (after compression) |

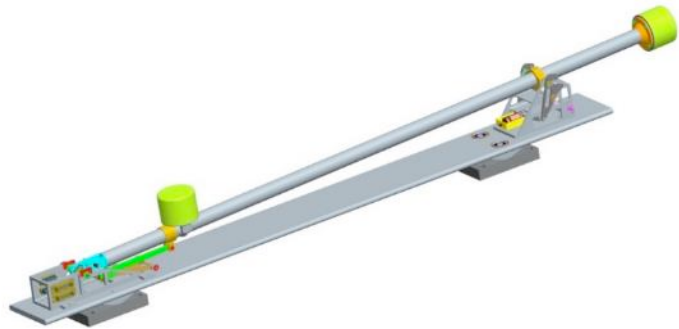

*Figure 8: Example of developed and tested boom with two flux-gate magnetometers for the KOMPSAT mission.*

Reaching a low noise and background in the magnetic field measurements is accomplished by several means. First, there are two booms in opposing directions of the spacecraft with the fluxgate sensors mounted at different distances to the body of the spacecraft. Since the spacecraft rotates at a spin rate of 15 rpm during the prime science phase, magnetic fields from the spacecraft can be separated from the to be measured magnetic fields of the environment based on the temporal signature. Moreover, since the spacecraft is of simple design accomplishing good levels of magnetic cleanliness is significantly easier than for a full-sized spacecraft. For example, on the probe there no reaction wheels for attitude control, just a cold gas system, and there are no solar panels.

The MAG sensor design, based on the heritage mentioned above, consists of two ring core elements of high-permeability material that serves as a concentrator for the external magnetic field. The excitation coils are wound tightly around the ring cores. A second set of coils - the three axes sensing coil system - picks up the induced signal. Additionally, a complete Helmholtz coil system is mounted within the sensor to provide feedback and to maintain the centre of the MAG sensor in a near-zero magnetic field. This feature keeps the sensor in a linear regime and avoids the need of range switching. Information about the ambient magnetic field is then extracted using both the sensed signal and the feedback current values. The sensor coil system is encapsulated in a cylindrical aluminium cap. Example of the sensor (MASCOT mission) is shown in Figure 9.

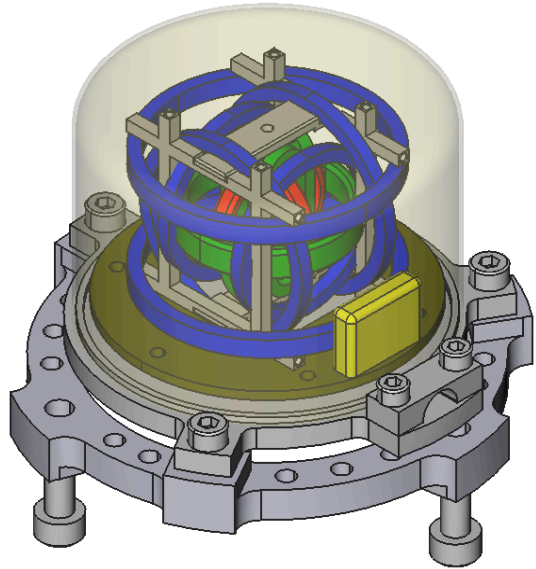

*Figure 9: Example of the MAG sensor, based on the MASCOT mission.*

The front-end electronics of MAG provides overall control of the magnetometer and communication with the main computer. The instrument is fully controlled by a single field-programmable gate array (FPGA), which provides functions for computation of the measured field and control over the feedback current, and handles the communication and switches the instrument into different operational modes. The sensor electronics generates an excitation AC current (fundamental excitation frequency of ~ 9.6 kHz). The front-end signal processing (synchronous detection and integration as well as the feedback value and the field calculation) is accomplished by logic blocks within the FPGA implemented as a RISC-like processor. The feedback field improves the overall linearity and stability of the magnetometer. It is supplied to all sensor elements via a set of two 16-bit digital-to-analogue converters (DACs) – called feedback DACs – and a separate pair of feedback coils (Helmholtz coils) per sensor axis. The sense signal from the analogue-to-digital converter (ADC) and the feedback values (setting of the feedback DACs) are used for calculating the magnetic field values.

Although the digital magnetometer concept requires an analogue-to-digital conversion at a high-data rate, it exhibits a number of advantages over the more traditional analogue fluxgate magnetometer. Early digitisation makes the sensed signal robust with respect to changes of the environmental temperature and supply voltages, as well as insensitive to electro-magnetic disturbances. Furthermore, no range switching is needed to achieve the required resolution, which reduces the complexity of design and data analysis.

For these types of magnetometers, a special analogue-to-digital and digital-to-analogue converter hybrid circuits have been developed. The ADC hybrid circuit implements one instrumentation amplifier, an 18 bit Megasample ADC, external adjustment of the amplification, and latch-up protection. The DAC hybrid implements two 16 bit DACs (coarse and fine ranges) and a current source for magnetometer feedback system. Additionally, both types provide a housing with additional spot shielding to enhance the radiation tolerance. The ADC hybrids have been used on the MASCOT mission. Both types, ADC and DAC, will be used for the ESA JUICE mission.

## 6.5 Radiation Monitor (RAD)

The radiation monitor (RAD) comprises: a telescope for electrons and protons, a heavy-ion monitor, a gamma ray monitor, a neutron monitor, a slow neutron counter, and an active shielding (see Figure 10). RAD will produce differential and selected integral fluxes for electrons, protons, gammas, and neutrons.

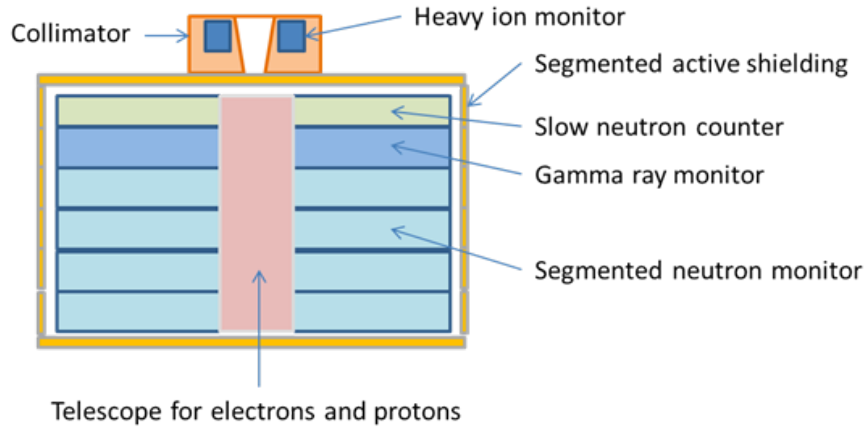

*Figure 10: Schematic drawing of EDP-RAD with its subunits.*

RAD builds on the heritage of the RADEM instrument for JUICE (Hajdas et al., 2015, 2016), with which it shares some of its capabilities and detection concepts; however, RAD differs from RADEM in a few key design aspects. First, it is equipped with additional capabilities specific to the EDP mission (Table 6**),** such as the gamma ray and neutron monitor. Second, it performs time resolved measurements while the probe is spinning during the descent with sufficient time resolution to distinguish the radiation coming from the surface of Europa from the one coming from the Jovian magnetosphere. Finally, it comprises an active shielding and coincidence detection schemes, and exploits mutual shielding between different parts of the instrument to minimise the passive shielding mass. The RAD detection systems combines the same mixed signal ASIC developed for RADEM with a new readout system, a fast 8-channels oscilloscope developed at PSI, based on the DRS4 chip.

*Table 6 RAD instrument capabilities.*

| Parameter | Value |
|---|---|
| **Electron Energy range** | 0.1 – 20 MeV* |
| **Proton energy range** | 5 – 250 MeV* |
| **Ion sensitivity** | He to O* |
| **Fast neutrons** | 5 – 100 MeV |
| **Thermal neutrons** | Sub eV |
| **Gamma Rays** | 0.5 – 2 MeV |

| **Peak electron flux** | $10^9$ #/(cm$^2$ s) (E > 100 keV) |
|---|---|
| **Accuracy - Sensitivity / Particle separation** | 10% |
| **Temporal resolution** | 1 s |
| **Telemetry data rate** | ≤ 100 bytes/s |
| **Volume** | ≤ 10 × 10× 10 cm$^3$ |

(*) = same values as RADEM for JUICE.

The electron and proton telescope utilises 16 Si-diodes separated to discriminate particles based on their characteristic energy loss in the absorber. The thickness of the absorber is optimised for the best possible detection of both electrons and protons in the energy range specified in Table 6. The detectors operate in coincidence, allowing for precise energy determination for protons. The electron telescope provides a set of unique response curves for energetic electrons as a function of energy, which is sufficient for the accurate unfolding of their energy spectra. Because numerous and energetic Jovian particles penetrating from the side of the instrument would diminish the telescope performance, the telescope structure is enclosed in a cylindrical copper shielding. Compared to RADEM, the thickness of RAD's copper shielding is reduced to necessary minimum by taking advantage of the shielding provided by the larger neutron and gamma ray detector box, and its anti-coincidence cast that surround the telescope. The only element incorporating a relevant volume of copper shielding is the entrance collimator: an appendix on top of the instrument that also contains the heavy-ion monitor. The electron and proton telescope allows for the determination of particle fluxes of ~$10^3$ cm$^{-2}$ s$^{-1}$, coming from the surface of Europa, with an accuracy of ~10% per 1-second measurement.

The heavy-ion monitor head is located inside the entrance cone of the electron and proton telescope to achieve a compact and mass efficient design. The detectors are arranged in a telescope and are enclosed into the body of the collimator to minimise side detections with larger energy depositions. Each layer of the telescope consists of several smaller Si-sensors connected in parallel to maximise the monitor area. The low-energy threshold for the heavy-ion sensor is set high enough not only to be above the noise level, but also to discriminate energy depositions from electrons and protons. A larger area and solid angle of the heavy-ion monitor allows for the determination of fluxes of heavy-ions with energy larger than a few hundred MeV with 10 % accuracy per 1-second measurement.

The gamma ray monitor consists of a scintillator for the identification of particles based on pulse shape analysis. The measurement is challenging because of the noise induced in the detector by the very-high rates and penetrating energies of the Jovian magnetosphere's electrons. For an efficient detection, a heavy scintillation material, such as CsI, is considered. CsI is a relatively fast scintillator that can operate together with an active shielding. Events registered simultaneously by the gamma ray monitor and the active shielding are rejected. Additional discrimination of penetrating electrons is provided by a slow-neutron counter, which is placed in front of the gamma ray monitor to take advantage of mutual radiation shielding. The active shielding is tailored to the size of the gamma ray monitor by using segmented plastic sheets that decrease the veto-time cadence of the system. Setting a low-energy threshold for the gamma ray monitor suppresses the bremsstrahlung photons from energetic electrons. Fast identification logic and pulse shape discrimination are followed up by a construction of rough energy spectra from the gamma ray counter, to meet the instrument energy range and telemetry data packet requirements. A 10% accuracy requirement is achieved for fluxes of ~100 cm$^{-2}$ s$^{-1}$.

The neutron monitor is based on stacked fast scintillators that are optically isolated and individually read out. To minimise the charged particle background, each segment has its own active shielding. This architecture provides a high efficiency and capability to identify neutral particles with a good signal to background ratio. The detection of gamma rays is suppressed by the low-energy detection threshold. In addition, the information from the gamma ray monitor allows for better analysis of the data and improved subtraction of photon events. As for the other detectors, the mutual shielding offered by the slow neutron counter and gamma ray monitor further suppresses the radiation background from both photons and charged particles, while only slightly modifying the incoming fast neutron flux. The readout logic is based on fast definition of the neutron signal, which followed by analysis of the analogue pulse-height provide coarse spectra of energy deposition. Preliminary estimations indicate that neutron fluxes above few-hundred cm$^{-2}$ s$^{-1}$ can be measured with an accuracy of ~10%.

The slow-neutron counter is based on a newly developed and tested ZnS:6LiF scintillator. Initial tests performed at PSI with the SINQ neutron source proved its high capability for slow neutron detection. The scintillator has virtually no gamma ray background, while the high amount of energy deposition by the slow neutrons allows to cut-off signals from charged particles. This detector is able to determine slow neutrons fluxes of ~100 cm$^{-2}$ s$^{-1}$.

For proper determination of neutral particle rates, an active shielding of segmented fast scintillators is applied. The active shielding is used to veto detections of the neutron and gamma ray monitors, whereas it is not used for the other detectors that have their own coincidence logic or low sensitivity for charged particles' energy deposition.

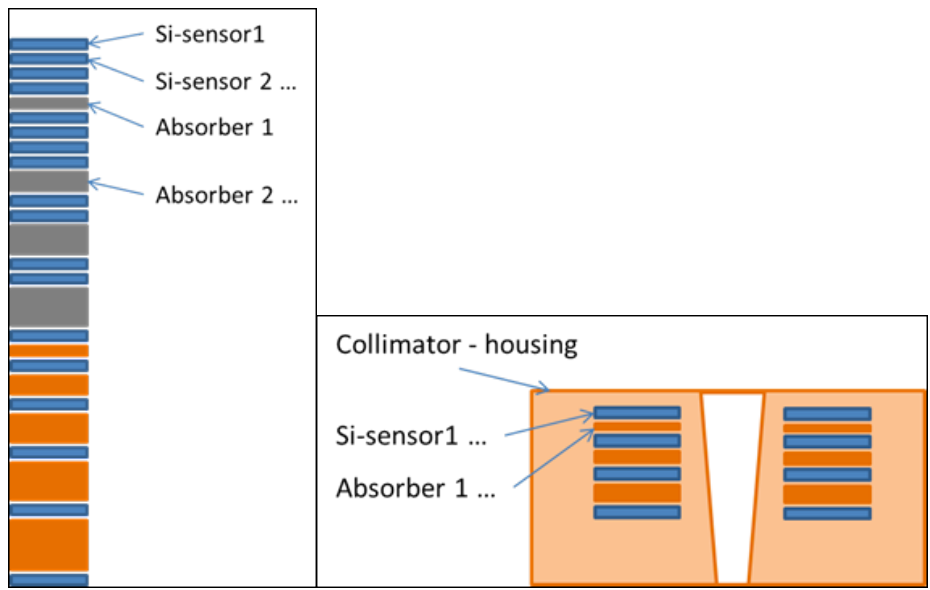

*Figure 11:* Schematic drawing of the telescope for electrons and protons (left) and of the heavy ion monitor (right).

# 7  Conclusions

Several missions took advantage of the combination of passing or orbiting spacecraft and small descent probes of various nature, especially when visiting remote or transient objects or the solar system, such as outer planets and their moons and comets. Successful missions that followed this approach include Rosetta (comet 67P/Churyumov–Gerasimenko orbiter and a cometary lander, Schulz et al. 2009), Cassini-Huygens (Saturn orbiter and a Titan atmospheric-entry probe, Matson et al. 2002; Lebreton and Matson, 2002), Galileo (Jupiter orbiter with an atmospheric-entry probe, Johnson et al. 1992), Deep Impact (Tempel 1 comet orbiter with a passive impactor, A'Hearn et al. 2005), LRO with the LCROSS impactor (Colaprete et al. 2010), and Chandrayaan-1 (Moon orbiter and impact probe, Goswami and Annadurai, 2009). In all cases, the science return from the synergetic operation of the mother spacecraft and the child probe was enormous, and often it was embedded in very few unique images or few minutes' worth of in situ data. Even though the NASA Europa Mission is designed for a very detailed investigation of Europa, it still would have been a great opportunity to add a descent probe and greatly amplify the science impact of an already very ambitious and capable science mission. Similar opportunities to augment the mission science by including a descent probe to a mission will arise in future mission, in particular in the outer solar system, where flybys at planetary objects without an atmosphere are foreseen.

The Europa Descent Probe represents a lightweight alternative option that provides a unique balance between science impact and engineering effort, in terms of complexity, risk, and Technology Readiness Level (TRL):

1.      The science performed by the EDP allows adding significant and valuable insights to the main science themes of the NASA Europa mission, and will additionally provide key engineering data that are necessary to plan future Europa landing mission.

2.      The EDP mission, even considering a 20% system margin, requires only about half of the available 250 kg mass budget, leaving room either for complementing the probe concept with additional science capabilities, or to accommodate other similarly lightweight piggy-back experiments on the mission.

3.      The EDP is based on standard and proven satellite bus technology and all its subsystems, including its science instruments, have a high TRL level (> 5). Moreover, many of the instruments would take considerable advantage of developments performed for ESA's JUICE mission.

4.      The EDP is compatible with the current mission profile of the NASA Europa mission, provided minor modifications could be implemented after a more detailed analysis of the trajectories. Certainly, the EDP does not require major changes, if any, of the Jovian tour of the main spacecraft, as it is compatible with the currently foreseen 3.9 km/s velocity of the early Europa's flybys (NASA Europa Study Team, 2012).

Although the study of EDP was conceived for an application for the Europa Mission mission, EDP can be easily adapted for other flyby situations.

•       Ganymede: since the Europa Mission encounters Ganymede multiple times at the beginning of the mission it offers in situ investigations near this moon at distances not covered by Europa Mission. However, since the JUICE mission will terminate its mission by crash-landing the spacecraft on the surface the additional science by a probe return will not be large.

- Callisto: is the moon for which this probe was originally conceived (Wurz et al. 2009, 2012a). Since neither the Europa Mission nor JUICE are getting closer than 200 km to the surface of Callisto the EDP can be employed there to its full scientific potential.

- Io: is an interesting but more complicated option since neither thev Europa Mission nor JUICE are flying by Io. Therefore, an EDP trajectory has to be found which uses its Europa flyby for an orbit-changing manoeuvre to bring EDP into a collision trajectory with Io. In addition, at least the communication system of EDP has to be augmented to cover the larger distance to the main spacecraft.

The concept of a descent probe deployed from a main spacecraft flying by the planetary body can be applied to virtually any atmosphere less planetary body of the solar system. In addition to the contribution of the science capabilities of the main mission, a descent probe contributes to engineering knowledge to be gained, for example when the body poses significant challenges for a landing scenario, either for the excessive engineering cost (e.g., mass) of a sophisticated lander, or for the unknown or harsh conditions at the surface which would make a landing mission too risky.